\documentstyle[prl,aps,epsf]{revtex}
\begin {document}

\draft

\twocolumn[\hsize\textwidth\columnwidth\hsize\csname@twocolumnfalse%
\endcsname

\title{Spin and chiral orderings of frustrated quantum spin chains}
\author{M. Kaburagi$^{1,3}$, H. Kawamura$^2$ and T. Hikihara$^3$}
\address{$^1$Faculty of Cross-Cultural Studies, 
Kobe University, Tsurukabuto, Nada, Kobe 657-8501, Japan\\
$^2$Faculty of Engineering and Design, 
Kyoto Institute of Technology, Sakyo-ku, Kyoto
 606-8585,
Japan\\
$^3$Graduate School of Science and Technology, Kobe University,
Rokkodai,  Kobe 657-8501, 
Japan}
\date{Draft: \today}
\maketitle
\begin{abstract}
Ordering of
frustrated $S$=$1/2$ and $1$ {\it XY\/} and Heisenberg spin chains 
with the competing nearest- and next-nearest-neighbor 
antiferromagnetic
couplings is studied by exact diagonalization 
and density-matrix renormalization-group methods. It is found that the
$S$=$1$ {\it XY\/} chain exhibits  both 
gapless and gapped `chiral' phases characterized
by the spontaneous breaking of parity, in which the
long-range order parameter is a chirality, 
$\kappa _i$=$S_i^xS_{i+1}^y-S_i^yS_{i+1}^x$, whereas 
the spin correlation decays either algebraically or 
exponentially. 
Such chiral phases are not realized  
in the $S$=$1/2$ {\it XY\/} chain
nor in the Heisenberg chains.
\end{abstract}
%\pacs{\rm PACS No: 75.10.Jm, 75.40.Cx}
]

Ordering of frustrated quantum spin chains has attracted 
considerable interest since these systems exhibit 
a rich variety of
magnetic phases due to the interplay between quantum effect and
frustration. 
We consider here an anisotropic frustrated quantum spin chain
described by the {\it XXZ\/} Hamiltonian,
\begin{equation}
   {\cal H} = \sum_{\rho=1}^{2}\Bigl\{J_{\rho}\sum_{\ell} 
       \bigl(S_\ell^x S_{\ell+\rho}^x + S_\ell^y S_{\ell+\rho}^y
           + \lambda S_\ell^z S_{\ell+\rho}^z\bigr)\Bigr\}~, 
\label{eq:Hami}
\end{equation}
where 
${\bf S}_{\ell}$ is the spin-$S$ 
spin operator at the $\ell$th site,
$J_{\rho}> 0$ is the antiferromagnetic interaction  
between the nearest-neighbor ($\rho$=$1$) and the 
next-nearest-neighbor ($\rho$=$2$) spin pairs, and 
$\lambda$ ($0 \le \lambda \le 1$) represents an exchange anisotropy.
Note that $\lambda\!=\!0$ and $\lambda\!=\!1$ correspond to the 
{\it XY\/} and Heisenberg chains, respectively.

The ground state phase diagram of the corresponding
$S$=$1/2$ system has been extensively studied either numerically
\cite{Tone-Hara,Oka-Nomu}
or analytically\cite{Maj-Gho,Haldane}. 
These studies have revealed that when $J_2$
is smaller than a critical value, {\it i.e.\/}, $j$$\equiv$$J_2/J_1
\leq j_c$,  the system is in the gapless spin-fluid phase 
in which the antiferromagnetic spin correlation decays 
algebraically.
By contrast, for larger values of $j>j_c$, the system is
in the dimer phase with a finite energy gap above the doubly
degenerate ground states. The dimer phase is
characterized by the spontaneously breaking of both parity and 
translation symmetries with preserving time-reversal symmetry.
The value of $j_c$ has been estimated to be
$j_c\cong \!$0.241 for the Heisenberg chain\cite{Oka-Nomu}.
% and 0.35 for the {\it XY\/} chain.
Although there is no
magnetic long-range order (LRO), 
the nature of the magnetic short-range order (SRO)
changes at the Lifshitz point $j_L$, where $j_L$$\simeq$$0.5$
for the Heisenberg chain\cite{Tone-Hara}.
For $j\leq j_L$, the system has
the standard N\'eel-type antiferromagnetic SRO and the structure
factor $S(q)$ has a maximum at $q=\pi $, while for $j>j_L$, 
the system has a helical SRO with the maximum of $S(q)$ at some
$q$=$Q<\pi $.

In the case of $S$=$1$, by contrast, 
no dimer phase occurs \cite{Tone-Kabu,Kol}. 
The Heisenberg chain is in the Haldane phase 
characterized by a singlet ground state and a finite energy
gap above it. A 
first-order transition takes place at $j$=$j_T\simeq 0.744$ between the
`single-chain' Haldane phase at $j<j_T$
and the `double-chain' Haldane phase at $j>j_T$\cite{Kol}.
In the {\it XY\/} case, on the other hand,  the situation remains
not entirely clear. Analytical studies based on the bosonization
method suggested that the 
gapless phase at $j=0$ (the so-called {\it XY\/}1 phase) 
extended to finite $j>0$\cite{Shima-Kubo} 
whereas numerical studies suggested
that the Haldane phase was stabilized for $j>0$\cite{Tone-Suzu}.
In any case, the fate of such
{\it XY\/}1 or Haldane phase at larger $j$ has not been 
clarified.

In the classical limit $S\rightarrow \infty $, the system exhibits a
magnetic LRO, either of the N\'eel-type ($j\leq 1/4$) or of the
helical-type characterized by the wavenumber
$q\!=\!\cos ^{-1}(-1/4j)$ ($j>1/4$). 
In the {\it XY\/} case, such helically ordered state 
possesses a twofold discrete degeneracy according as
the helix is either right- or left-handed, in addition to a
continuous degeneracy associated with the original $U(1)$
symmetry of the {\it XY\/} spin. This discrete degeneracy
is characterized by mutually opposite
signs of the total chirality defined by\cite{Kawa1}
\begin{eqnarray}
\kappa &=& \frac {1}{N}\sum _i\kappa _i,
\label{eq:chi}
\\
\kappa _i &=& S_i^xS_{i+1}^y-S_i^yS_{i+1}^x=[{\bf S}_i\times 
{\bf S}_{i+1}]_z, \nonumber
\end{eqnarray}
where $N$ is the total number of spins. 
Chirality is invariant under both $U(1)$ spin-rotation and 
time-reversal operations, but changes its sign under parity
operation. In the Heisenberg case, while there no longer exists a
discrete  chiral degeneracy, one can still formally define 
the chirality by Eq.(\ref{eq:chi}) as a $z$-component of 
the vector chirality, 
{\bf $\kappa $}$_i$={\bf S}$_i\times ${\bf S}$_{i+1}$. 
Note that the chirality defined above  is distinct
from the scalar chirality of the Heisenberg spin 
often discussed in the literature\cite{Frahm}
defined by
$\chi _i={\bf S}_{i-1}\cdot {\bf S}_i\times {\bf S}_{i+1}$: The
scalar chirality of the Heisenberg spin changes sign under the
time-reversal operation unlike the chirality (2).
Since the classical chain (1) always has a planar spin order,
the scalar chirality $\chi _i$ vanishes trivially.

Recent studies on various frustrated {\it classical\/} systems
with continuous symmetry 
have revealed that such chiral degrees of freedom often give rise to
novel ordering behaviors such as  phase transitions of
new universality class\cite{Kawa2}, 
novel magnetic phase diagrams with 
new multicritical behavior\cite{Kawa2}, 
and even a novel `chiral phase' 
in which only the chirality exhibits a LRO without the standard
spin LRO\cite{triXY,ffXY,XYSG}. 
Meanwhile, systematic studies of the possible chiral order
in frustrated {\it quantum\/} spin chains has been scarce so far.  

In this Letter, we perform a numerical study of 
the ground-state properties of
a class of $S$=$1/2$ and $1$ frustrated spin chains (1)
in search for a possible
chiral order. We have found that both gapless and gapped 
chiral phases, where
the chirality has a finite LRO while the spin correlation
falls either algebraically or exponentially, are realized 
for a wide range of $j$ 
in the $S$=$1$ {\it XY\/} chain,
but not in the $S$=$1/2$ {\it XY\/} chain
nor in the $S$=$1/2$ and $1$ Heisenberg chains.

In order to probe the possible chiral order, we first 
calculate the square of the
total chirality $<\kappa ^2>$, where the chiral order parameter 
is defined by Eq.(2) now for 
%\newline

\begin{figure}[ht]
\begin{center}
\noindent
\epsfxsize=0.44\textwidth
%\epsfile{file=fig1a.ps,scale=0.36}
\epsfbox{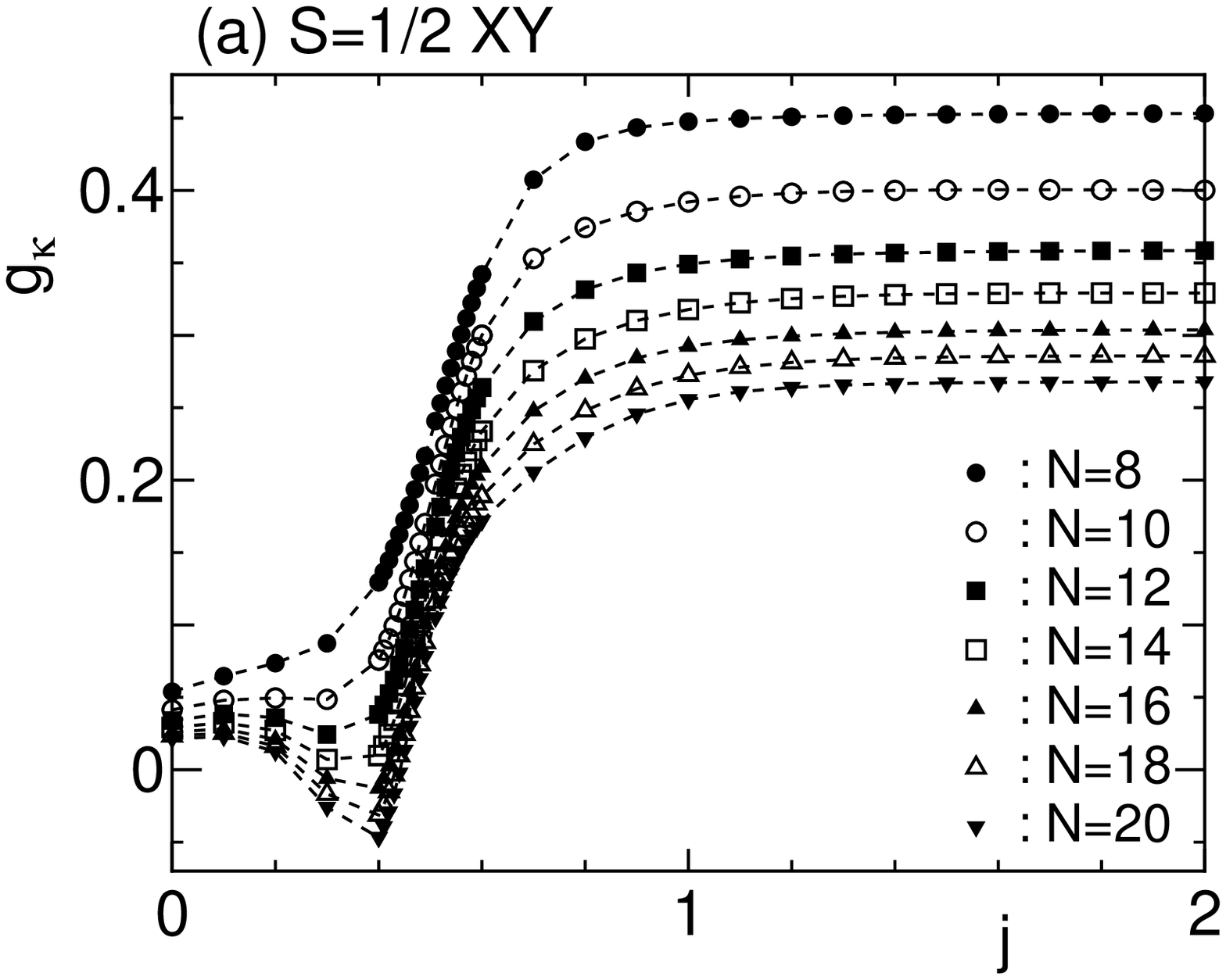}
\end{center}
\vspace{-0.5cm}
\begin{center}
\noindent
\epsfxsize=0.44\textwidth
%\epsfile{file=fig1b.ps,scale=0.36}
\epsfbox{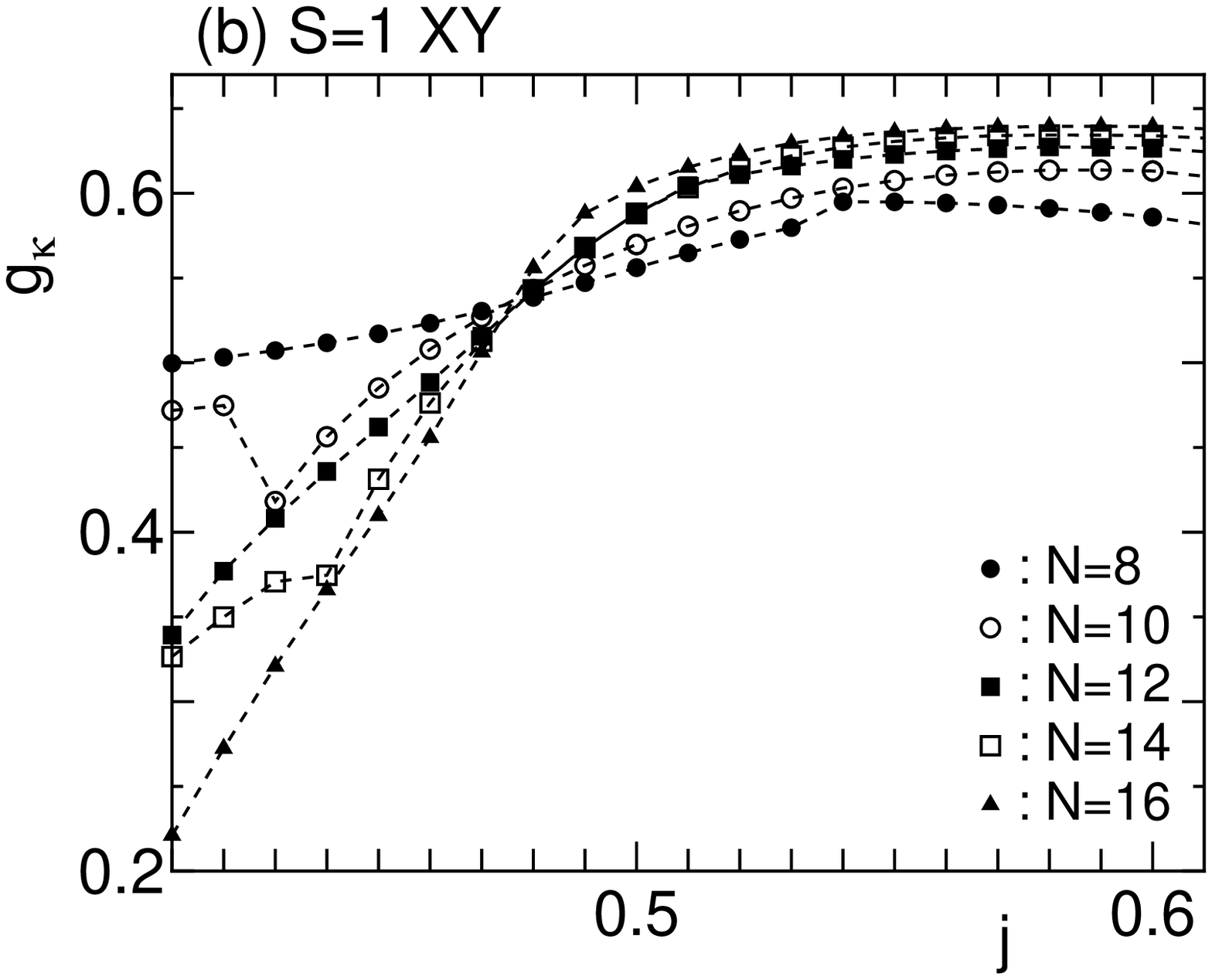}
\end{center}
\caption{Binder parameter of the chirality
$g_\kappa$ versus $j$: 
(a)  $S$=$1/2$ {\it XY\/}chain; 
(b)  $S$=$1$ {\it XY\/}chain.
}
\end{figure}

\noindent
quantum spins, together with 
the associated Binder parameter,
\begin{equation}
g_\kappa =(1-\frac{<\kappa ^4>}{3<\kappa ^2>^2}),
\label{eq:gchi}
\end{equation}
by exactly diagonalizing finite open chains with even $N$ 
up to $N$=$20$ ($S$=$1/2$) or $N$=$16$ ($S$=$1$)\cite{Rem1}. 
All ground states are found to belong
to the subspace of $S_{\rm total}^z=0$ with
even parity.

The calculated Binder parameters of the chirality $g_\kappa $
of the $XY$ chain  
are shown  in Fig.1 for various $N$ as a function of $j$. As shown
in Fig.1(a), $g_\kappa $ of the $S=1/2$  chain 
constantly decreases
with increasing $N$ for any $j$, demonstrating the absence of
chiral order. In contrast, $g_\kappa $ of
the $S=1$  chain exhibits a significantly different
behavior. As can be seen from Fig.1(b), 
$g_\kappa $  for various $N$ cross at almost the same $j$,
$j=0.475\pm 0.005$, above which $g_\kappa $ increases 
with increasing $N$, 
indicating the existence of a finite chiral LRO for larger $j$.

In the case of Heisenberg chains, on the other hand, we have
found that for both cases of $S=1/2$ and $1$ the calculated
$g_\kappa $ (not shown here) constantly decreases with increasing
$N$, which clearly shows that
the chiral order is not realized in the Heisenberg chains.

In order to examine more closely the nature of the transition 
and of the possible phases,
we  calculate, 
concentrating on the $S$=$1$ {\it XY\/} open chain, 
the two-point chiral, spin, and string correlation functions defined 
by
\begin{eqnarray}
C_\kappa (r) &=& <\kappa_{r_0-r/2}\kappa_{r_0+r/2}>, \label{eq:cchi} \\
C_{s}(r) &=& <S_{r_0-r/2}^x S_{r_0+r/2}^x>,
\label{eq:cspin} \\
C_{\rm str}(r) &=& 
<S_{r_0-r/2}^z \bigl( \exp i\pi \sum_{j=r_0-r/2+1}^{r_0+r/2-1} S_j^z \bigr) 
S_{r_0+r/2}^z>,
\label{eq:cstring} 
\end{eqnarray}
by means of the density-matrix renormalization-group 
%(DMRG) 
method.\cite{White}.
Here $r_0$ represents the center position of  open chain, 
{\it i.e.,\/} 
$r_0$=$N/2$ for even $r$ and 
$r_0$=$(N+1)/2$ for odd $r$. 
We employ the infinite system method by using 
$M$ block states ($M\leq 300$) 
in the subspace of $S_{\rm total}^z$=$0$ with even parity.
Convergence of the results with respect to $M$ 
has been checked by consecutively increasing $M$. 
In the chiral phase, the chiral correlation function $C_\kappa (r)$
tends to a finite constant at large $r$, while in the Haldane phase,
the string correlation function $C_{str}(r)$ tends to a finite
constant. 

The calculated $r$-dependence of the chiral, string, and spin 
correlation functions 
are shown in Fig.2(a)-(c) on log-log plots
for several typical values of $j$.
As can be seen from Fig.2(a), the data of $C_\kappa (r)$
for $j>j_{c1}\simeq 0.473$ bend up at larger $r$ suggesting a finite
chiral LRO, while they bend down for  $j<j_{c1}$ suggesting 
the absence of chiral order. Our 
estimate of $j_{c1}=0.473\pm 0.001$ is consistent with the 
estimate from the Binder parameter given above.  
Meanwhile, as shown in Fig.2(b), the data of $C_{str} (r)$ 
for $j<j_{c2}\simeq 0.490$ 
bend up for larger $r$ suggesting a finite
string order characteristic of the Haldane state,
while for $j>j_{c2}$ they show a linear behavior signaling 
a power-law decay of the string correlation.
The existence of the Haldane phase for smaller $j$ is consistent with
the previous finding of Ref.\cite{Tone-Suzu}.

An interesting observation here is that our estimate of 
$j_{c2}=0.490^{+0.010}_{-0.005}$
is distinctly larger than that of $j_{c1}=0.473\pm 0.001$,
which means that {\it there exist two different types of chiral 
phases\/},
one with a finite string order ($j_{c1}<j<j_{c2}$) 
and the other without the string order ($j>j_{c2}$).
This finding is also supported by the behavior of the spin
correlation function $C_s(r)$ in Fig.2(c), in which $C_s(r)$
divided by the oscillating factor $\cos (Qr)$ is shown. 
Indeed, the data of $C_s(r)$
exhibit a linear behavior for $j>j_{c2}$ indicating a power-law decay
of helical spin correlations (gapless state), 
while for $j<j_{c2}$ they bend down suggesting an exponential-decay
of helical spin correlations (gapped state).
Note that above the Lifshitz point, which we estimate to be
$j_L=0.313\pm 0.001$
for the $S=1$ {\it XY\/} chain,
the system exhibits a helical SRO
characterized by a wavevector $Q<\pi $. On the other hand, 
the absence of magnetic (spin) LRO has rigorously been proven 
for any $j$ and for general $S<\infty $\cite{Momoi}.
The existence of a novel intermediate phase, a
gapped chiral phase, 
in a narrow but finite range of $j$ can clearly be seen in
Fig.2 from the behavior of the correlation functions  at
$j=0.477$ which lies between $j_{c1}$ and $j_{c2}$.

Thus,  on increasing $j$, 
the $S=1$ {\it XY\/} spin chain undergoes two successive
transitions, first  at $j=j_{c1}$ from the Haldane phase
with no chiral order to the gapped chiral phase (or the chiral
Haldane phase),
then  at $j=j_{c2}$ from the gapped chiral phase
to the gapless chiral phase.
In the gapped chiral phase, the chiral and  string LRO coexist
with exponentially-decaying  spin correlations, whereas in the
gapless chiral phase, only the chirality shows a LRO with
algebraically-decaying spin and string correlations.  
In the gapped (gapless) chiral phase, the ground state is doubly
degenerate, each of which is characterized by the opposite sense
of the chirality, with (without) a finite gap above it.

The gapped chiral phase is characterized by a spontaneously broken
parity with preserving both translation and time-reversal
symmetries.
From a broken symmetry, the transition at $j=j_{c1}$ 
is expected to be of the Ising-type. We  extract the chiral
correlation length $\xi _\kappa $ from the calculated $C_\kappa (r)$
and fit it to the standard power-law form 
$\xi_{\kappa} \sim (j_{\rm c1}-j)^{-\nu_{\kappa}}$. Our present 
estimate,
$\nu _\kappa \simeq 0.9\pm 0.1$ appears to be 
slightly smaller than, but not inconsistent with
the 2D Ising value $\nu _\kappa =1$. 
Further detailed study of the
critical properties, including the nature of
the KT-like transition at $j=j_{c2}$, is in progress.

It should be noticed that Nersesyan {\it et. al.\/} recently
discussed 
for the $S$=$1/2$ {\it XY\/} chain in the limit 
of large $j$ the possibility of the parity breaking 
\cite{Ne}.  
In contrast to this suggestion, for the $S$=$1/2$ {\it XY\/} and
Heisenberg 
\newpage

\begin{figure}[ht]
\begin{center}
\noindent
\epsfxsize=0.45\textwidth
%\epsfile{file=figchl.ps,scale=0.36}
\epsfbox{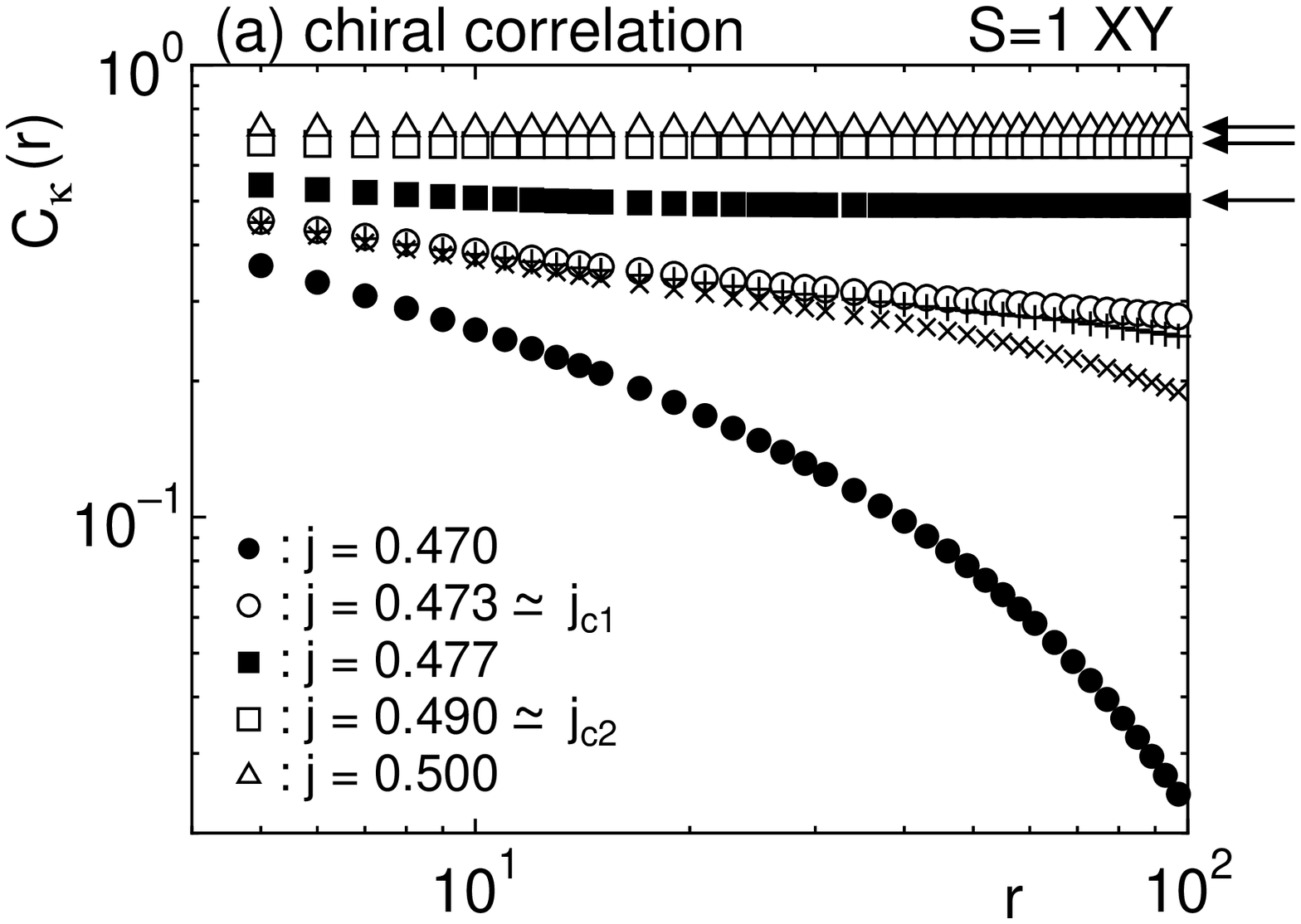}
\end{center}
\end{figure}

\vspace{-1.0cm}
\begin{figure}
\begin{center}
\noindent
\epsfxsize=0.45\textwidth
%\epsfile{file=fig1a.ps,scale=0.36}
\epsfbox{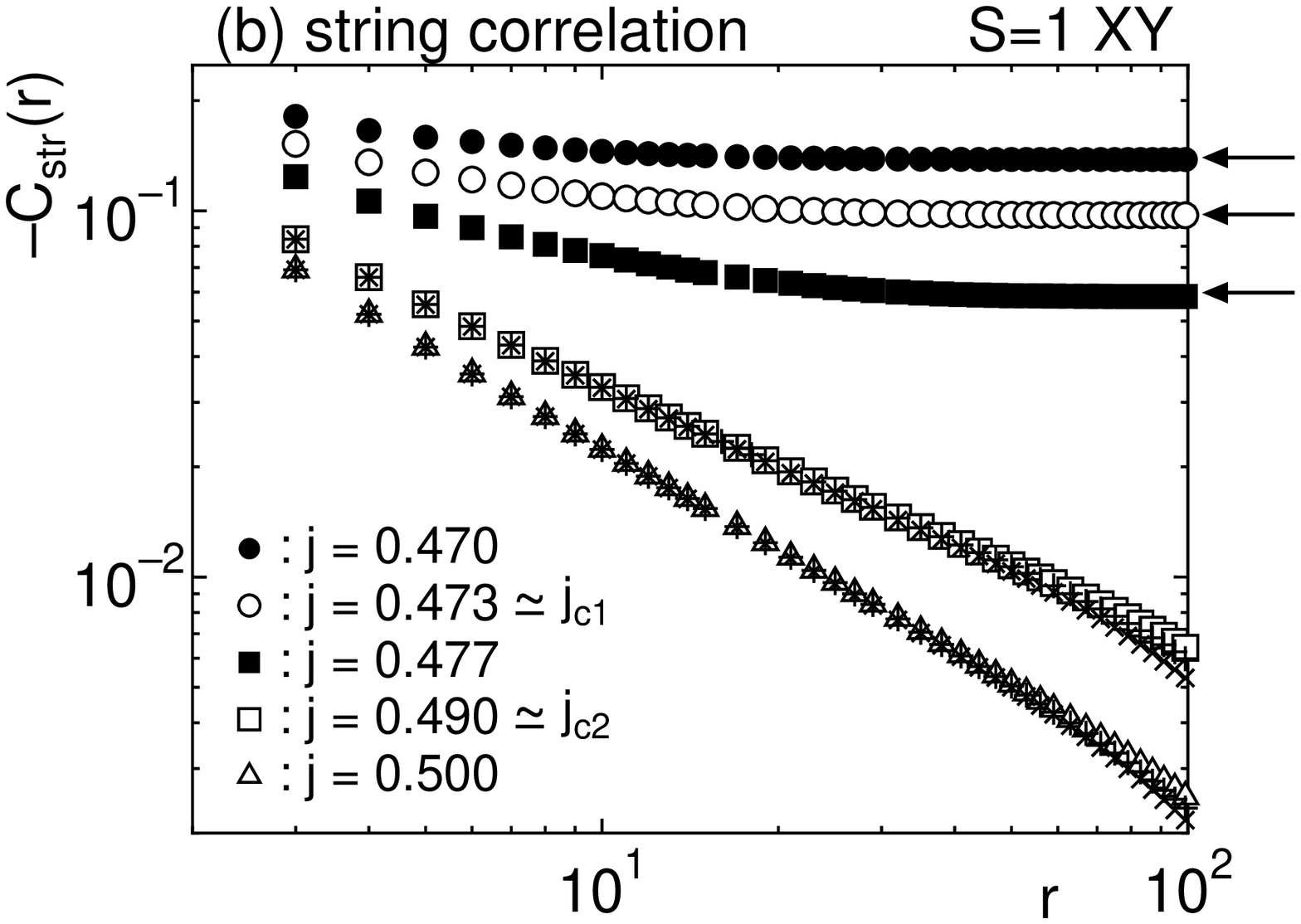}
\end{center}
\end{figure}

\vspace{-1.0cm}
\begin{figure}
\begin{center}
\noindent
\epsfxsize=0.43\textwidth
%\epsfile{file=figchl.ps,scale=0.36}
\epsfbox{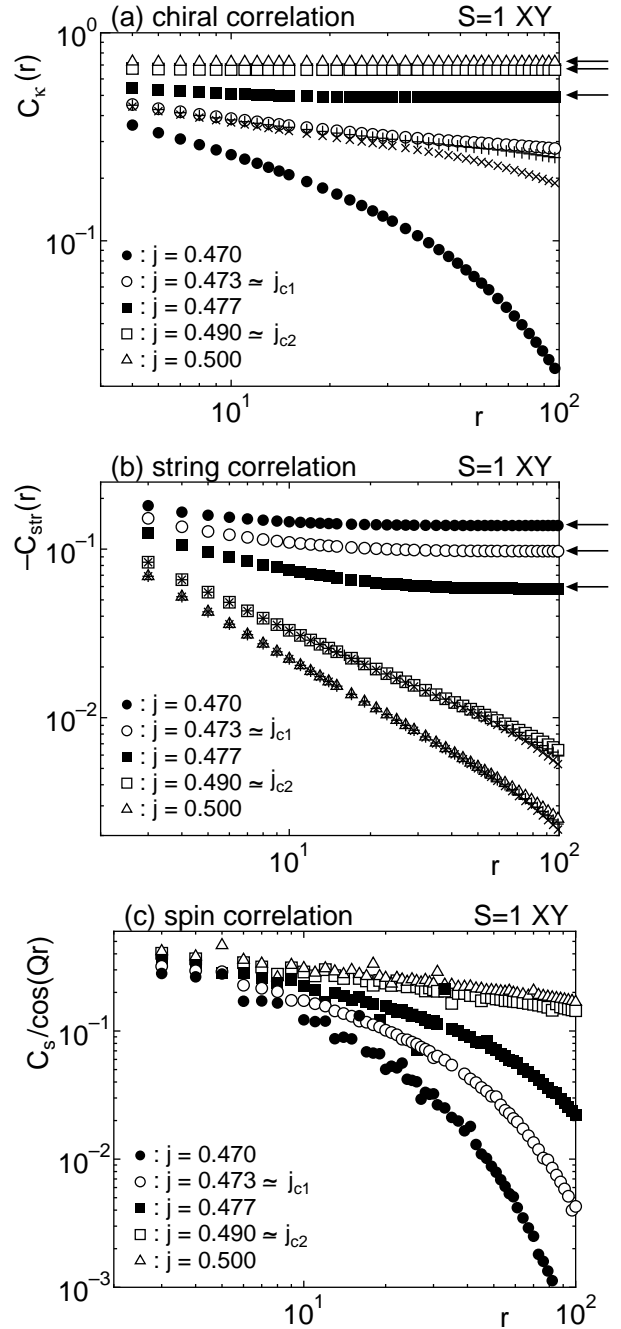}
\end{center}
\caption{Correlation functions versus $r$ on log-log plots
for various $j$: 
(a) chiral correlation $C_\kappa (r)$;
(b) string correlation $-C_{\rm str}(r)$;
(c) spin correlation $C_{s}(r)$ divided by the oscillating factor
$\cos (Qr)$. Arrows in the figures represent the extrapolated
$r=\infty $ values. The number of block states is equal to $M=300$.
To illustrate the $M$-dependence, we also
indicate by crosses the data for $M=220$ and $260$ for several cases 
where the $M$-dependence 
is relative large.
Note that some of the data points for larger $r$ are  omitted for
clarity.
%Rather large scatter of the cata of $C_s(r)$ in (c)
%is due to ...........
}
\end{figure}

\noindent
chains,
we did not find the chiral phase \cite{comment}
nor any new phase except for the
well-known spin-fluid and dimer phases, corroborating the
type of the phase diagram  previously reported by Haldane
\cite{Haldane} and by Tonegawa {\it et al\/} \cite{Tone-Hara}.
The chiral phase found here
appears to be specific to the $S=1$ {\it XY\/} chain.
Such tendency may roughly be understood by noting that only
the XY chain sustains an Ising-like discrete chirality which 
has a stronger ordering tendency than the continuous spin variable
while
quantum fluctuations might be strong enough in the $S=1/2$ chain 
to wash out even the chiral ordering.

It might also be interesting to notice 
that the observed novel transition behavior of
the $S=1$ {\it XY\/} chain has a similarity to that of the
2D frustrated classical {\it XY\/} models such as the 
triangular-lattice {\it XY\/} antiferromagnet\cite{triXY} or the 
Josephson-junction array in a magnetic field\cite{ffXY}. 
In these classical
systems, Miyashita and Shiba, and more recently Olsson observed
by Monte Carlo simulations 
that the thermal phase transitions occurred 
in two steps with two types
of chiral phases, each characterized by 
exponentially-decaying and algebraically-decaying
spin correlations. At the moment, we donot know whether there
exists a deeper connection between the two systems. 

Finally, we wish to briefly discuss the possible
experimental implication of our
results. In order to observe the chiral phase, one needs
to prepare an $S=1$ {\it XY\/} zig-zag chain with its $j$ value in
a suitable range. In the presence of weak 3D interchain
interaction, while the gapless chiral phase stabilized 
for larger $j$ is
expected to show the standard helical spin LRO,   
the gapped chiral phase could remain gapped. Hence, it is 
challenging to experimentally observe 
the gapped chiral state in an appropriate
model material. One problem here might be that the gapped chiral 
phase is realized in a rather
narrow range of $j$. It might then be necessary to tune the $j$ value
by some experimental method, such as by applying pressure.
Once an appropriate sample could be prepared, it is in principle 
possible to measure the chirality by using, {\it e.g.\/}, polarized
neutrons.\cite{Kawa2,Plumer}

In summary, from numerical studies of the ground-state properties of
a class of $S$=$1/2$ and $1$ frustrated spin chains
we have found that both gapless and gapped 
chiral phases, where
the chirality has a finite LRO while the spin correlation
falls either algebraically or exponentially, are realized 
for a wide range of $j$ 
in the $S$=$1$ {\it XY\/} chain, but 
not in the $S$=$1/2$ {\it XY\/} chain
nor in the $S$=$1/2$ and $1$ Heisenberg chains.
Further details of the results including  
the critical properties and the full phase diagrams 
will be reported elsewhere. 

We thank Prof. T. Tonegawa  for valuable discussion.
Numerical calculations were carried out in part at the Yukawa
Institute Computer Facility, Kyoto University.


\begin{thebibliography}{10}
\bibitem{Rem1}
More precisely, we calculate $g_\kappa$ of  open chains
by using 
$\kappa= \frac {1}{N-1}\sum_{i=1}^{N-1}\kappa _i$.
\bibitem{comment}
For example, we calculated $g_\kappa $ by exact diagonalization
up to $j=1000$, and observed a behavior similar to the one shown in
Fig.1(a).
\end{thebibliography}
\end{document}